\documentclass{PoS}
\title{Nuclear Masses in Astrophysics}
\ShortTitle{Nuclear Masses in Astrophysics}
\author{Christine Weber$^a$, \speaker{Klaus Blaum}$^{b}$, and Hendrik
Schatz$^{c}$\\
\llap{$^a$}Department of Physics, University of Jyv\"askyl\"a, FI-40014 Jyv\"askyl\"a, Finland\\
\llap{$^b$}Max-Planck-Institut f\"ur Kernphysik, D-69117 Heidelberg, Germany\\
\llap{$^c$}NSCL, Michigan State University, East Lansing, MI 48824-1321, USA\\
E-mail: \email{christine.weber@jyu.fi},
\email{klaus.blaum@mpi-hd.mpg.de}, \email{schatz@nscl.msu.edu}}
\abstract{Among all nuclear ground-state properties, atomic masses
are highly specific for each particular combination of $N$ and $Z$
and the data obtained apply to a variety of physics topics. One of
the most crucial questions to be addressed in mass spectrometry of
unstable radionuclides is the one of understanding the processes
of element formation in the Universe. To this end, accurate atomic
mass values of a large number of exotic nuclei participating in
nucleosynthesis are among the key input data in large-scale
reaction network calculations. In this paper, a review on the
latest achievements in mass spectrometry for nuclear astrophysics
is given.}
\FullConference{10th Symposium on Nuclei in the Cosmos\\
         July 27 - August 1 2008\\
         Mackinac Island, Michigan, USA}
\begin{document}
\vspace{-2.5mm}
\section{Introduction}
\vspace{-1mm}To date mainly two experimental approaches for
high-precision mass measurements on short-lived nuclides with the
required relative mass uncertainties of $10^{-7}$ exist, namely
storage ring and Penning trap mass spectrometry \cite{Blau2006}.
The installation of such ion-storage devices at on-line facilities
allows to combine the advantages of extremely sensitive,
high-precision experiments for studies on a vast number of exotic
nuclides that are readily provided today. Beams far off the valley
of $\beta$ stability are available at accelerator facilities such
as ISOLDE/CERN, GSI in Darmstadt, JYFL in Jyv\"askyl\"a, or the
NSCL at Michigan State University. They are either produced in
proton-induced spallation, fragmentation, or fission reactions in
a thick target (ISOL approach) or via fusion-evaporation or
fragmentation and a subsequent in-flight separation. Depending on
the particular production scheme, several initial steps including
laser- or surface-ionization techniques \cite{Kost2002} in the
ISOL approach, or fast stopping and extraction schemes are
employed \cite{Ayst2001,Boll2005}.

Whereas neutron-deficient nuclides close to the $N = Z$ line and
in the vicinity of the rapid proton capture ($rp$)-process are
routinely produced and measurements even beyond the proton drip
line have been successfully conducted farther up around $A = 145$
\cite{Raut2008}, neutron-rich nuclides studied are located close
to the $N = 50$ shell closure \cite{Haka2008,Baru2008} or span
from the nickel ($A = 73$) to the palladium ($A = 120$) isotopes
\cite{Raha2007a,Raha2007b,Dela2006,Hage2007a,Hage2007b}. However,
the presumed path of the $r$-process is merely reached by high
precision mass measurements in the region around $N = 50$ and
close to $N = 82$ within the tin isotopes \cite{Dwor2008}.
$^{130}$Cd has been reached with the less precise $\beta$-endpoint
method \cite{Dill2003}.

\vspace{-2.5mm}
\section{Mass spectrometry of short-lived isotopes}
\vspace{-1mm}
\subsection{Penning traps}
\vspace{-0.5mm}At present six Penning trap facilities are
operational at accelerators. These experimental setups include
ISOLTRAP \cite{Mukh2007} at ISOLDE/CERN, the Canadian Penning trap
(CPT) \cite{Sava2006} at the Argonne National Laboratory ANL,
SHIPTRAP \cite{Bloc2005} at GSI/Darmstadt, the JYFLTRAP facility
\cite{Kolh2004} in Jyv\"askyl\"a, LEBIT \cite{Ring2006} at
NSCL(MSU), and TITAN \cite{Dill2006} at TRIUMF/Vancouver. The
first Penning trap mass spectrometer at a nuclear reactor is
presently in the commissioning phase at TRIGA-Mainz
\cite{Kete2008}. The common pre-requisite in all experiments is an
ion deceleration, accumulation, and cooling as successfully
obtained in linear radiofrequency quadrupole (rfq) coolers and
bunchers \cite{Niem2001,Herf2001,Schw2003} with the goal to
deliver ion pulses with a well-defined emittance to the subsequent
Penning trap setups.

\begin{figure}[h] \center
\includegraphics[width=0.43\textwidth]{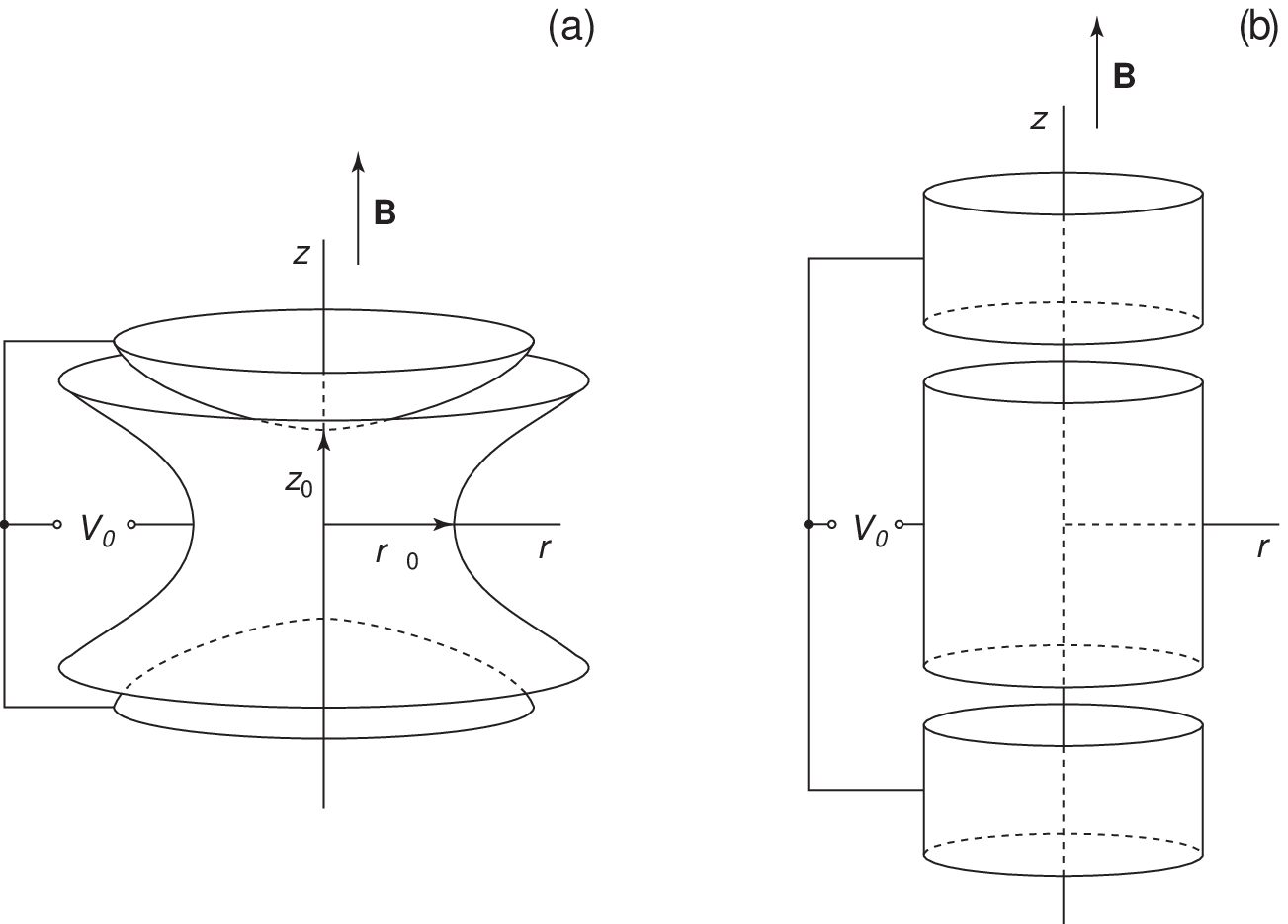}
\hspace{0.02\textwidth}
\includegraphics[width=0.35\textwidth]{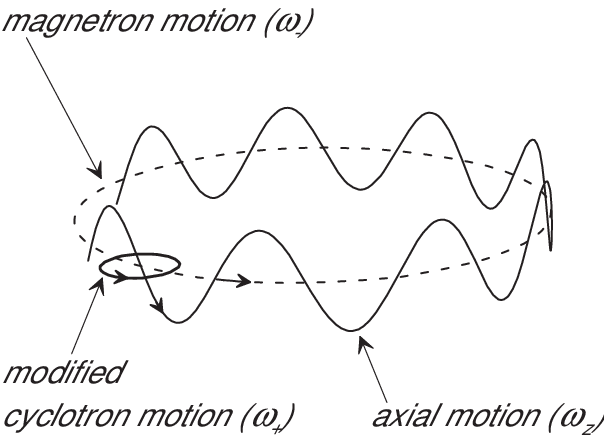}
\caption{Left: Penning trap configurations. A trap consists of a
central ring electrode and two endcap electrodes. A hyperbolical
electrode geometry (a) generates the quadrupolar shape of the
electrostatic storage potential. Traps with a cylindrical
electrode geometry (b) are frequently used due to easier capture
properties. Right: Schematic of the characteristic eigenmotions of
a charged particle in a Penning trap.} \label{motion}
\end{figure}

Figure \ref{motion} shows two geometric configurations of a
Penning trap as employed in mass spectrometry on short-lived
nuclides. It is operated in a combination of a high-field ($B \sim
7~\mbox{T}$), high-homogeneity ($\Delta B/B \le
10^{-6}/\mbox{cm}^{3}$) superconducting magnet and a quadrupolar
electrostatic storage potential that is applied as a voltage
difference $V_0$ between the ring electrode and both endcaps of
the trap (see Fig. \ref{motion}). The resulting ion motions in the
combined fields are depicted in the right part of the figure. They
consist of a superposition of three ideally independent harmonic
oscillations with characteristic eigenfrequencies $\omega$: an
axial motion ($\omega_z$) and two radial motions, the modified
cyclotron motion ($\omega_+$) and a slow $\vec{E} \times \vec{B}$
drift, the magnetron motion ($\omega_-$). Their frequencies are
\begin{equation}
    \label{2-6}
    \omega_z = \sqrt{\frac{qV_0}{md^2}}
    \hspace{2em} \mbox{and} \hspace{2em}
    \omega_\pm = \frac{\omega_c}{2} \pm \sqrt{\frac{\omega_c^2}{4} -
    \frac{\omega_z^2}{2}},
\end{equation}
where $\omega_c = \frac{q}{m}B$ is the free cyclotron frequency
and $d$ the characteristic trap dimension \cite{Brow1986}. The sum
frequency of both radial motions in an ideal Penning trap
$\omega_+ + \omega_- = qB/m$ is directly related to the
charge-to-mass ratio $q/m$. It is determined in a time-of-flight
detection method \cite{Graf1980}. Here, ions are released from the
trap and their flight time to a detector in the magnet's fringe
field is recorded. Prior to their ejection, ions are excited with
an azimuthal quadrupolar RF field around $\omega_c$
\cite{Koni1995a}. In case of a resonant excitation at $\omega_c$
the ion's radial kinetic energy is maximized, which is transformed
into an additional axial acceleration while ions travel along the
magnetic field gradient.
Atomic masses $m$ are derived using alternating calibration
measurements of nuclides with precisely known mass values or
close-by carbon cluster ions $^{12}\mbox{C}_n^+$ for absolute mass
calibration \cite{Blau2002,Kell2003}. In this way, relative mass
uncertainties $\delta m/m$ as low as few times $10^{-8}$ are
routinely achieved.

%
\vspace{-2mm}
\subsection{Ion storage rings}
\begin{figure}
\center \vspace{-0.2mm}
\includegraphics[angle=-90,width=0.49\textwidth]{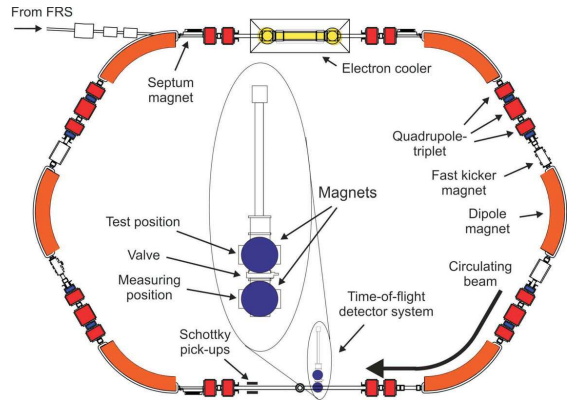}
\caption{(Color online) Layout of the Experimental Storage Ring
(ESR) at GSI/Darmstadt. The circumference of the ring is
$108~\mbox{m}$. Ions are injected from the Fragment Separator
(FRS). In experiments with the Schottky technique, ions are cooled
in the electron cooler and their relative velocity spread is
reduced by five orders of magnitude. Induced signals of revolving
ions are detected at the Schottky pick-up electrodes. Experiments
in the isochronous operation mode register the revolution times of
circulating ions in a dedicated time-of-flight detector system.}
\label{esr}
\end{figure}

\vspace{-0.5mm}A large-scale approach for mass spectroscopy of
stored ions are direct measurements in storage rings. At present,
such measurements are routinely performed at the GSI Experimental
Storage Ring ESR \cite{Fran2008}, and are being started at the
CSRe storage ring in IMP Lanzhou \cite{Meng2008}. Mass
measurements are also planned at RIKEN with a new RI-RING project
\cite{Yama2008}. At GSI, exotic nuclides are produced in
projectile-fragmentation reactions or fission in thin targets,
selected in the Fragment Separator (FRS) and are subsequently
injected into the ESR. Figure \ref{esr} shows a schematic layout
of the ESR storage ring. For ions circulating in the ring the
relative difference in the mass-to-charge ratio $m/q$ of the
revolving ion species is expressed as
\begin{equation}
\label{revol}
                        \frac{\Delta f}{f}
                    = - \frac{1}{\gamma_t^2} \frac{\Delta (m/q)}{m/q} +
                    \left(1 - \frac{\gamma^2}{\gamma_t^2}\right) \frac{\Delta v}{v},
\end{equation}
where $\Delta f/f$ is the relative difference in the revolution
frequency, $\gamma$ is the Lorentz factor, and $v$ is the
velocity.

Two different approaches exist for a direct determination of
$m/q$: for rather long-lived nuclides with half-lives of more than
a few seconds ions can be cooled in the electron cooler of the
ring with the aim to reduce their relative velocity spread
($\Delta v/v$) to few $10^{-7}$. Hence, the term on the right in
Eq. (\ref{revol}) is negligible and the revolution frequencies
directly correspond to $m/q$. Here, induced signals are recorded
in the Schottky pick-up electrodes and are subsequently Fourier
analyzed. A multitude of ion species is stored simultaneously in
the ring and hence, the masses of several nuclides can be
determined in an extremely efficient way, whereas those ions with
precisely known masses are used for calibration. Since multiply
charged ions are detected, single-ion sensitivity is reached in
these measurements. Moreover, the identification of ions in
different charge states is useful as a check for internal
consistency of the obtained data set. Typical relative mass
uncertainties obtained are few times $10^{-7}$, corresponding to
$\delta m \approx 30~\mbox{keV}$ for $A \approx 90 - 200$
\cite{Litv2005}.

Alternatively, the so-called isochronous operation mode of the
storage ring is employed \cite{Haus2000}. Here, the Lorentz factor
$\gamma$ is selected to be equal to the transition energy
$\gamma_t$ and again, the second term in Eq. (\ref{revol})
vanishes such that revolution frequencies are proportional to
$m/q$. This method does not require electron cooling and is
well-suited for nuclides with very short half-lives as low as a
few ten $\mu\mbox{s}$ \cite{Stad2004}. In this case a dedicated
time-of-flight detector is used to measure the revolution times of
stored ions \cite{Trot1992}. In recent studies the masses of
several neutron-rich fission fragments were determined using an
additional $B\rho$ determination prior to the injection into the
ring and typical mass uncertainties of about $\delta m \approx
120~\mbox{keV}$ were obtained \cite{Sun2008a}.

\vspace{-2.5mm}
\section{Neutron-deficient nuclei and implications for $rp$- and $\nu p$-process nucleosynthesis}
\vspace{-1mm}A large set of new data on neutron-deficient nuclei
relevant for studies of the astrophysical $rp$-process
\cite{Wallace.Woosley:1981,Scha1998} and the recently proposed
$\nu p$-process \cite{Froh2006,Prue2006} were obtained with
Penning trap mass spectrometers world-wide. They support the
modelling of these nucleosynthesis pathways, which aims at
understanding the final elemental abundances and energy
production, and at comparing model results with the growing number
of astronomical observations in a quantitative way. The
$rp$-process in type~I x-ray bursts \cite{Scha1998} starts, for
example, from a breakout of the hot CNO cycle by a sequence of
$\alpha$-induced reactions and proceeds then via rapid
proton-capture reactions and subsequent $\beta^+$ decays close to
$N = Z$. In particularly hydrogen rich bursts it can reach up to
tellurium, where the predominant $\alpha$ instability,
$^{107}\mbox{Te}$ and $^{108}\mbox{Te}$, returns the flow into the
closed SnSbTe cycles \cite{Scha2001}.

In addition to $\beta$-decay half-lives, precise atomic mass data
are among the most critical nuclear parameters in reaction network
calculations for nucleosynthesis. The detailed reaction flow is
determined by individual mass differences, the single-proton
separation energies, moreover mass values are used to calculate
reaction energies to model capture processes. To this end, mass
uncertainties $\delta m$ on the order of less than
$10~\mbox{keV}$, {\it i.e.} $\delta m/m \le 10^{-7}$, are required
\cite{Scha2006}, which are routinely achieved in precision Penning
trap mass spectrometry.

In the lower part of the $rp$-process, direct mass measurements of
the bare short-lived $^{44}$V, $^{48}$Mn, $^{41}$Ti, and $^{45}$Cr
ions where obtained in isochronous mass spectrometry (IMS) at the
ESR storage ring \cite{Stad2004}. The mass uncertainties of few
100~keV obtained in these studies contribute with first data on
proton separation energies beyond scandium and the $rp$-process
flow through the titanium to manganese isotopes is discussed.

Along the $rp$-path, the so-called waiting point nuclei,
$^{64}\mbox{Ge}$, $^{68}\mbox{Se}$, $^{72}\mbox{Kr}$, and
$^{76}\mbox{Sr}$, are of particular importance, since they can
cause a delay in the process. At a waiting point, a
($p,\gamma$)-($\gamma,p$) equilibrium is reached between
proton-capture and photodisintegration reactions while further
proton captures are hindered by low $Q$ values to the next,
proton-unbound isotone. The process slows down until these
nuclides undergo $\beta$ decay. The effective lifetime of a
waiting point can be as long as its $\beta$-decay half-life,
modified by the lifetime against possible two-proton captures
\cite{Scha2006}. Studies of these nuclei and their close vicinity
are needed to constrain the limits on the effective lifetimes,
which depend exponentially on the $Q$ value for proton capture.
Some of the critical masses around $^{64}\mbox{Ge}$ and
$^{68}\mbox{Se}$ have been both determined with CPT
\cite{Clar2007,Clar2004} and LEBIT \cite{Schu2007}, respectively,
reducing the uncertainties in the effective lifetimes.
Measurements on $^{72-74}\mbox{Kr}$ were obtained with ISOLTRAP
\cite{Rodr2004}. The latter study indicates that the $\beta$-decay
lifetime is modified by less than 20\%. In addition, several
nuclides such as neutron-deficient strontium isotopes up to the
waiting point $^{76}\mbox{Sr}$ \cite{Sikl2005}, or
neutron-deficient isotopes of selenium and bromine \cite{Herf}
were studied at ISOLTRAP.

\begin{figure}
\vspace{0.395cm} \center
\includegraphics[width=0.73\textwidth]{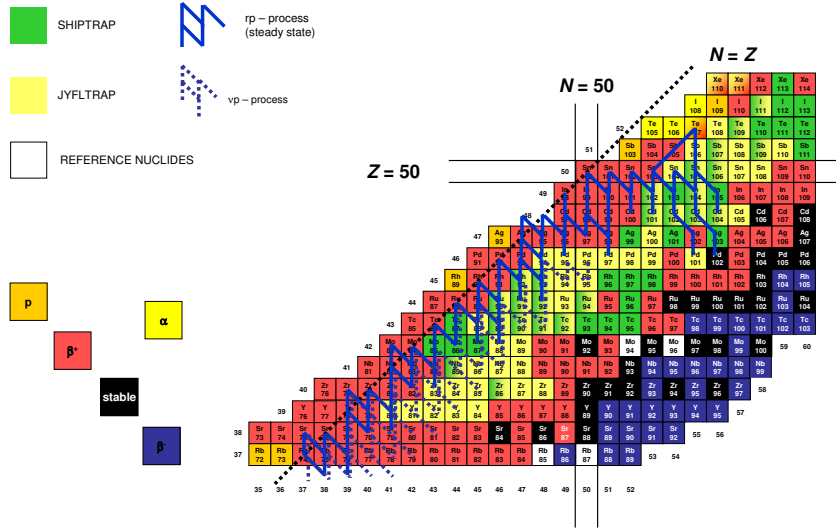}\vspace{-0.2cm}
\caption{(Color online) Section of the nuclear chart with the
upper region of the $rp$- and $\nu p$-process pathways. Nuclides
studied with the JYFLTRAP (yellow) or the SHIPTRAP (green) mass
spectrometer are indicated. For a detailed review see text.}
\label{chart}
\end{figure}

Across the higher-mass region of the $rp$-process path above $A =
80$ broad mass determinations have been performed with JYFLTRAP,
SHIPTRAP and CPT. Figure \ref{chart} gives an overview on all
nuclides studied in either of the first two experiments. A
possible pathway of the $rp$-process for steady-state burning
(from Schatz \cite{Scha2001}, solid lines) is shown together with
a possible path of the $\nu p$-process (dashed lines). The data
amount to about 75 newly determined mass values which are not
included in the published issue of the Atomic Mass Evaluation 2003
\cite{Audi2003a}. Their mass uncertainties $\delta m$ are well
below the wanted limit of $10~\mbox{keV}$ as required in
nucleosynthesis calculations \cite{Scha2006}.

Masses of yttrium, zirconium, and niobium isotopes were determined
at JYFLTRAP employing heavy-ion induced $^{32}\mbox{S} +
(^{54}\mbox{Fe}~\mbox{or}~^{\mbox{\scriptsize{nat}}}\mbox{Ni})$
reactions \cite{Kank2006} with resulting typical uncertainties of
$7~\mbox{keV}$. In a common approach of JYFLTRAP and SHIPTRAP the
masses of 21 further nuclides up to and including $^{96}\mbox{Pd}$
were studied \cite{Webe2008}, and almost half of them were
determined experimentally for the first time. The results of both
Penning trap experiments are in excellent agreement with each
other, and a weighted mean with an improved final uncertainty as
low as $2.9~\mbox{keV}$ was given. Results for the most exotic
isotopes deviate substantially from data in AME2003, which are
mostly stemming from $\beta$-endpoint measurements and
extrapolations of systematic trends. The impact of the new results
was studied in nucleosynthesis calculations of the $\nu
p$-process. Detailed reaction flow pattern were compared with
calculations that only include the data of AME2003. Since the new
mass values, for example the one of $^{88}\mbox{Tc}$ results in a
proton-separation energy which is $1~\mbox{MeV}$ smaller than in
the AME2003 systematics, the reaction flow around $^{88}\mbox{Tc}$
is strongly modified. However, the final abundances for the $\nu
p$-process calculations were found to be almost unchanged.

Nuclides above $^{96}\mbox{Pd}$ were investigated in independent
experiments at either SHIPTRAP \cite{Mart2007} or JYFLTRAP
\cite{Elom2008}. Whereas heavy-ion induced fusion reactions of
$^{58}\mbox{Cr}$ and $^{58}\mbox{Ni}$ beams on $^{58}\mbox{Ni}$
targets were employed for measurements at SHIPTRAP, measurements
in Jyv\"askyl\"a focused on the verification of production
possibilities of these nuclides in light-ion induced reactions
with $p$ or $^{3}\mbox{He}$ beams. Among the nuclides studied only
the most neutron-deficient ones are relevant for $rp$-process
nucleosynthesis. Here, discrepancies of up to $2\sigma$ are
observed in the cadmium isotopes $^{101,102,104}\mbox{Cd}$.
However, this apparent discrepancy can possibly be solved by new
data from ISOLTRAP along the cadmium isotopic chain
$^{99-109}\mbox{Cd}$ \cite{Brei2009}. Most recently, the mass
values of the exotic $^{85}\mbox{Mo}$ and $^{87}\mbox{Tc}$
nuclides were determined for the first time in experiments at
SHIPTRAP and the mass values of $^{86,87}\mbox{Mo}$,
$^{93-95}\mbox{Tc}$, $^{94,96}\mbox{Ru}$, and $^{96-98}\mbox{Rh}$
were measured and substantially improved \cite{Bloc2008}.

The mass values of several nuclides from molybdenum to rhodium
were determined with the CPT \cite{Clar2005,Fall2008a,Fall2008b}.
As the production of the light $p$ nuclei, $^{92,94}\mbox{Mo}$ and
$^{96,98}\mbox{Ru}$ is not quantitatively understood, {\it i.e.}
these nuclides are observed in the Solar System in greater
abundance than predicted in $p$-process theory, the latter
publications focus on the inconsistency in the molybdenum
isotopes. The $\nu p$-process was introduced to explain the
creation of these nuclides in proton-rich ejecta of supernova
explosions in neutrino-driven winds \cite{Froh2006,Prue2006}. It
was shown that to reproduce the observed abundance ratios, current
models would require a separation energy $S_p$ of
$^{93}\mbox{Rh}$, calculated as $- M(^{93}\mbox{Rh}) +
M(^{92}\mbox{Ru}) + M(^{1}\mbox{H})$, of $1.64 \pm 0.1~\mbox{MeV}$
\cite{Fisk2007,Hoff2008}. However, all Penning trap experiments
and the estimation in AME \cite{Audi2003a} yield a separation
energy that differs considerably from this value. This discrepancy
might indicate either the presence of a different production site,
or deficiencies in the astrophysical models. Though there is an
agreement in the $S_p$ value between the trap data, the mass
values of both nuclides involved, $^{93}\mbox{Rh}$ and
$^{92}\mbox{Ru}$, deviate by about 1.5$\sigma$ in measurements
from SHIPTRAP, JYFLTRAP \cite{Webe2008} and the ones of the
Canadian Penning trap \cite{Fall2008a}.

In the endpoint region of the $rp$-process the mass values of
thirteen nuclides were determined at JYFLTRAP (see Fig.
\ref{chart}). The resulting one-proton separation energies in the
studied antimony isotopes are particularly relevant in order to
estimate the fraction of proton captures that flow into the closed
SnSbTe cycles. With a new value of the proton separation energy
$S_p$ for $^{105}\mbox{Sb}$, determined indirectly from the
$\alpha$ energy of $^{109}\mbox{I}$ in Ref. \cite{Mazz2007}, the
formation of a significant cycle at $^{104}\mbox{Sn}$ is excluded.
The new data in this region, in particular the first experimental
determinations on $^{106}\mbox{Sb}$, $^{108}\mbox{Sb}$, and
$^{110}\mbox{Sb}$ test the concept of the $rp$ endpoint and
explore a possible leakage past the closed SnSbTe cycles \cite{JYFL2009}.\\
\vspace{-2.5mm}
\section{Neutron-rich nuclei and implications for $r$-process nucleosynthesis}\vspace{-1mm}
The rapid neutron-capture process ($r$-process) is responsible for
the origin of about half of the heavy elements beyond germanium in
the cosmos. Elements such as europium, gold, platinum, or uranium
are mainly produced in the $r$-process \cite{Cowa1991,Pfei2001}.
However, where this process occurs is not known with certainty and
its understanding is one of the greatest challenges of modern
nuclear astrophysics. The isotopic abundance pattern not only
depends on the chosen astrophysical environment, {\it e.g.} the
neutron density, but also sensitively on the underlying nuclear
physics processes and parameters \cite{Krat1993,Wana2004}. Thus,
accurate mass data on the extremely neutron-rich nuclei
participating in the $r$-process are of utmost importance to
compare the signature of specific models with astronomical data
now becoming available from observations of metal-poor stars in
the Galaxy \cite{Trur2002}.

Recently, the TOF-B$\rho$ technique that includes a position
measurement for magnetic rigidity correction has been implemented
at the NSCL facility using the A1900 separator and the S800
spectrograph. The first experiment, focused on the neutron-rich
isotopes in the region of Z $\sim$ 20-30, important for r-process
calculations as well as for calculations of processes occurring in
the crust of accreting neutron stars, has been successfully
performed \cite{Estr2008,Mato2008}.

\begin{figure}
\center
\includegraphics[width=0.43\textwidth]{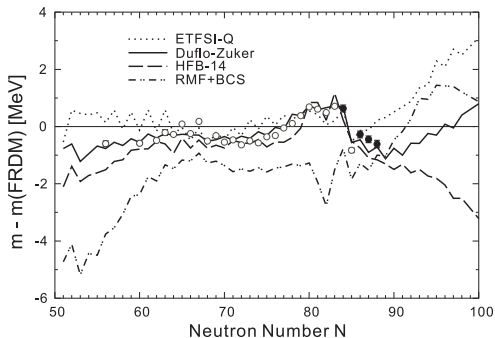}
\caption{Mass comparison of model predictions and experimental
data for iodine isotopes (from Ref. \cite{Sun2008a}).}\label{Z53}
\end{figure}

A number of neutron-rich nuclei close to the $r$-process path have
been studied with the IMS technique at the ESR \cite{Sun2008a}.
The experimental mass values for iodine isotopes including the
newly determined ones are compared to predictions of several
modern theories in  Fig. \ref{Z53}. It is clear to see that the
ETFSI-Q model has a divergent trend beyond $N = 80$ and fails to
describe the new mass data. The models in this figure span from
macroscopic-microscopic to self-consistent microscopic models,
which predict very different structure of $r$-process nuclides.
Calculations of reaction pathways based on these models yield
abundances which sometimes differ by several orders of magnitude
\cite{Sun2008b}. It is therefore essential to provide the
experimental basis for testing these predictions and consequently
improving the underlying models.

Most modern $r$-process models predict that the $r$-process occurs
at high temperatures of a billion degrees or more. At such high
temperatures energetic photons can excite nuclei in a way that
they emit neutrons. These so-called photodisintegration reactions
can counteract the rapid neutron captures. At which nucleus this
happens within an isotopic chain depends mainly on the binding
energy of the neutrons in the nuclei and therefore on the nuclear
masses. In this context $r$-process waiting points, {\it i.e.}
nuclides where photodisintegration wins over neutron capture and
thus the $r$-process temporarily stops and waits for the $\beta$
decay into the next isotopic chain, are of highest importance and
their masses can be provided with relative uncertainties as low as
$10^{-7}$. Nuclear masses therefore largely determine the path of
the $r$-process on the chart of nuclides. Together with the
$\beta$-decay half-lives of the waiting point nuclei, the masses
determine also the speed of the process and the final abundance
pattern. The most influential waiting points in the $r$-process
path are believed to be $^{80}$Zn, $^{130}$Cd, and $^{195}$Tm,
responsible for producing the pronounced abundance peaks observed
around mass numbers $A$ equals 80, 130, and 195, respectively (see
Ref. \cite{Pfei2001}).

To date, $Q$-value measurements related to these waiting points
were only performed using the $\beta$-endpoint technique in the
case of $^{130}$Cd with a mass uncertainty of 150~keV
\cite{Dill2003} and very recently using precision Penning trap
mass spectrometry in the case of $^{80}$Zn with an uncertainty of
only a few~keV \cite{Haka2008,Baru2008}. While the precision
achieved with the $\beta$-endpoint technique is not sufficient to
perform reliable $r$-process calculations for particular key
nuclides, the extremely low uncertainty meanwhile routinely
achievable with Penning trap mass measurements allowed, in the
case of $^{80}$Zn, to extract a well-defined map of conditions for
a major $r$-process waiting point to be on the reaction path.
\vspace{-2.5mm}
\section{Conclusions and future perspectives}\vspace{-1mm}
By the huge amount of improved mass data available, a more
reliable modelling of the astrophysical nucleosynthesis processes
is feasible. New network calculations should ideally comprise
up-to-date data from different experiments to provide best
possible mass values including extrapolations. Together with
better defined conditions within the actual astrophysical sites,
these enable to answer detailed questions on both sides of the
valley of $\beta$ stability.

Unfortunately, most of the extremely neutron-rich nuclei in the
$r$-process are still beyond the reach of existing Nuclear Physics
accelerator facilities since they have very small production
rates. However, upgrades toward future radioactive beam (RIB)
facilities such as the future facility FAIR at GSI in Darmstadt,
Germany \cite{Henn2004}, the BigRIPS separator at RIKEN/Japan
\cite{Kubo2007} or a new advanced rare-isotope accelerator in the
US \cite{xxx}, aim at higher beam intensities and the production
of neutron-rich nuclides farther away from stability and may
overcome this limitation. For example, the future ILIMA@FAIR
project will enable to address nuclei with production rates as low
as 1 ion per day/week, corresponding to production rates on the
order of picobarn \cite{Sun2008a}.

The authors highly appreciate the detailed comments to the
manuscript from Yuri A. Litvinov. Financial support by the HGF
(contract VH-NG-037), by the German BMBF (contract 06MZ215), by
the EU within NIPNET (contract HPRI-CT-2001-50034), Ion Catcher
(contract HPRI-CT-2001-50022), EURONS (JRAs TRAPSPEC and DLEP),
the Academy of Finland under the Centre of Excellence Programmes
2000-2005 (Nuclear and Condensed Matter Physics Programme),
2006-2011 (Nuclear and Accelerator Based Physics Programme at
JYFL), and NSF grant numbers PHY-0606007 and PHY-02-16783 (Joint
Institute for Nuclear Astrophysics) is gratefully acknowledged.

%

\end{document}